\documentclass[5p,number,times]{elsarticle} 

\usepackage{graphicx}
\usepackage{amsmath}
\usepackage{amssymb}
\usepackage{epsf,latexsym}
\usepackage{color}
\newcommand{\ket}[1]{\ensuremath{\left| #1 \right\rangle}}

\newcommand{\IPR}[3] {\ensuremath{\left\langle {#1} {\left| {#2} \right|} {#3}\right\rangle}}
\newcommand{\EX}[1] {\ensuremath{\left\langle #1 \right\rangle}}
\newcommand{\be}{\begin{equation}}
\newcommand{\bel}[1]{\begin{equation}\label{#1}}
\newcommand{\ee}{\end{equation}}
\newcommand{\ba}{\begin{eqnarray}}
\newcommand{\ea}{\end{eqnarray}}

\newcommand{\OP}[2] {\ensuremath{\left| {#1} \right\rangle \left\langle{#2}\right|}}

\newcommand{\Fig}[1]{Fig.~\ref{#1}}
\newcommand{\Eq}[1]{Eq.~(\ref{#1})}

\newcommand{\NEWT}[1]{\textcolor{black}{#1}}
\newcommand{\NEW}[1]{\textcolor{black}{#1}}
\newcommand{\TIM}[1]{\textcolor{black}{#1}}

\newcommand{\BUE}{Centre for Theoretical Physics, %
             The British University in Egypt, %
             Postal No. 11837, P.O. Box 43, Egypt.}

\newcommand{\LBRO}{Department of Physics, Loughborough University, %
              Loughborough, Leics LE11 3TU, United Kingdom}
\newcommand{\LEEDS}{Quantum Information Science, School of Physics and Astronomy, %
University of Leeds, Leeds LS2 9JT, UK}
\newcommand{\NII}{National Institute of Informatics, 2-1-2 Hitotsubashi, Chiyoda-ku, Tokyo 101-8430, Japan}
\newcommand{\NTT}{NTT Basic Research Laboratories, 3-1, Morinosato Wakamiya Atsugi-shi, Kanagawa 243-0198, Japan}

\journal{Physics Letters A}

\begin{document}


\begin{frontmatter}

\title{Quantum measurement with chaotic apparatus}
\author[lu,bue]{M. J. Everitt}
\ead{m.j.everitt@physics.org}
\author[nii,ntt]{W. J. Munro}
\author[leeds]{T. P. Spiller}
\address[lu]{\LBRO}
\address[bue]{\BUE}
\address[nii]{\NII}
\address[ntt]{\NTT}
\address[leeds]{\LEEDS}

\date{\today}

\begin{abstract}
We study a dissipative quantum mechanical model of the projective measurement of a qubit. 
We demonstrate how a correspondence limit, damped quantum oscillator can realise chaotic-like or periodic trajectories that emerge in sympathy with the projection of the qubit state, providing a model of the measurement process. \end{abstract}


\end{frontmatter}

\section{Introduction}

Given its general nature, plethora of applications and ability to explain forces and the structure of matter, quantum mechanics is considered to be amongst the most successful of physical theories. Furthermore, the majority of today's advanced technologies and their future growth, depends on quantum mechanics---indirectly if not directly.  Intuitively one might therefore expect that everything about quantum mechanics would be well understood. Surprisingly, this is not the case. Not only does a unique interpretation of the theory prove elusive, with individuals able to express preferences ranging from the Copenhagen interpretation, many-worlds to objective reality; even fundamentals such as the \emph{``measurement''} of \emph{``observable''} quantities are not fully understood. Indeed the debate as to exactly what is meant by the measurement of a quantum object has continued since Born's paper in 1926~\cite{Born:1926p1789} until the present day (see, for example, \cite{wheeler83,joos85,zurek91,bell90,Schlosshauer::2004p1267,bha2,Jacobs:2006p159,Leggett:2008p1513,Wis96,Bhattacharya:2000p1198,Greenbaum:2007p432,Pen98,Everitt:2005p95}). In this work we illustrate how a classical record of a quantum measurement can arise. 

We propose a mechanism for the realisation of projective
measurements simply from the framework of
Schr\"odinger evolution, modified to allow for an environment.
Our model comprises a qubit coupled to an open
quantum mechanical measurement device which is arranged to operate in
its classical (correspondence) limit. This is achieved by
ensuring that the classical action of the measurement device
is large (and remains so) with respect to a Planck cell, and introducing
a dissipative environment to suppress quantum interference
effects in this device (a paradigm that is well studied and known
to be rather effective -- see, for example,~\cite{Schlosshauer::2004p1267,Per98,Brun:1996p393,Habib:2006p1182,Hab98,Schack:1995p389,Bhattacharya:2000p1198,Greenbaum:2007p432,Kapulkin:2008p1005,Everitt:2009p1694}).

\NEW{We note that there have been many different models of the measurement process that usually involve introducing some kind of environmental decoherence (such as in quantum trajectories methods). In these approaches the central strategy is to model the process of measurement as the effect of external degrees of freedom, as an environment, acting on the quantum object and infer the measurement from this. 
In this work we model the measurement device itself and introduce the environment coupled to this device, simply to achieve its classical limit via decoherence. 
Here the measurement device comprises a single oscillator that in its classical limit has only two degrees of freedom (position and momentum). Furthermore, in its correspondence limit this oscillator can exhibit dissipative chaos -- such non-linear dynamical behaviour is not realisable by the Schr\"odinger equation for the isolated oscillator. Consequently, our system is not Hamiltonian chaotic nor is it related to quantum chaos in terms of random matrix theory. In this case the realisation of chaotic like trajectories is a clear signature of classicality (see, for example, ~\cite{Schlosshauer::2004p1267,Per98,Brun:1996p393,Habib:2006p1182,Hab98,Schack:1995p389,Bhattacharya:2000p1198,Greenbaum:2007p432,Kapulkin:2008p1005,Everitt:2009p1694}).
Hence, our work is different in that the environment is just a tool to render the measurement device classical. We do not infer any associated measurement, this emerges from the dynamics.}

\section{The Model}

Our strategy is to find a suitable system comprising a (quantum) measurement device and a quantum object, such that (i) the former projects the later into some state, (ii)  the measurement device has a discernibly different characteristic classical like behaviour depending on the measured state and (iii) for a \emph{``good''} measurement the Born rule holds. Put simply, the probabilities of projecting the quantum object follow from the appropriate squared amplitudes in its initial state. We therefore study a system of two components, the first being a quantum oscillator that has a well defined correspondence limit and whose classical like dynamics is manifestly different depending on the quantum state of the other component of the system. The second component should be a simple quantum device, such as a qubit, arranged so that there is no direct coupling between its possible states. Our model Hamiltonian that satisfies this requirement is a driven non-linear oscillator:
\bel{eq:Ham0}
H=\frac{3}{4}p^2+\frac{\beta^2}{4}q^4-\frac{1}{4}q^2+\frac{g}{\beta}\cos(t)q-H_\mathrm{int}.
\ee
As we elect to measure the $z$-eigenvalue of the qubit we take a cross-Kerr like interaction\TIM{, that couples the oscillator to a degenerate qubit,} of the form
\bel{eq:Hint}
H_\mathrm{int}=\frac{1}{4}\left(p^2+q^2\right)\sigma_z=\frac{1}{2} \left(a^{\dagger} a + \frac{1}{2}\right)\sigma_z.
\ee
\begin{figure}
\begin{center}
\resizebox*{0.4\textwidth}{!}{\includegraphics{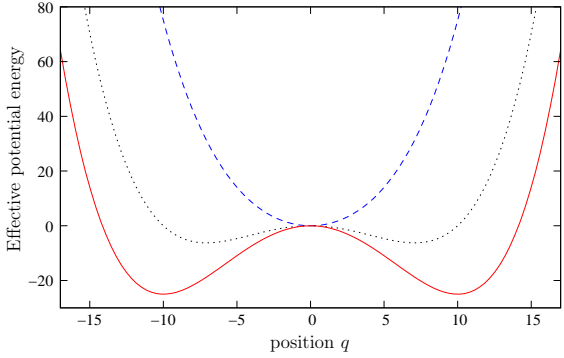}}
\end{center} \caption{(colour online) Effective potential for the Hamiltonian of \Eq{eq:Ham} where $\beta=0.1$ and (a) $\EX{\sigma_z}=1$ solid (red) line (b)  $\EX{\sigma_z}=-1$ dashed (blue) line and (c)  $\EX{\sigma_z}=0$ dotted line.
}
\label{fig:pot}
\end{figure}
Here $q$ and $p$ are the dimensionless position and conjugate momentum for the oscillator mode (i.e. defined such that the annihilation operator $a=(q+ip)/\sqrt{2}$) and $\sigma_z$ is the usual Pauli operator for the qubit. 

\TIM{A cross-Kerr like interaction is a physically reasonable one as it is well known~\cite{plk} to emerge as the dispersive (far-detuned) limit of the Jaynes-Cummings (JC) interaction Hamiltonian~\cite{Jaynes:1963p1700}. The JC interaction between a qubit and an oscillator field is widely applicable, to systems such as Rydberg atoms~\cite{[iii]}, NMR studies of nuclei~\cite{[iv],[v]}, Cooper pair boxes ~\cite{[vi]}, Cavity Quantum Electrodynamics~\cite{[vii]} and trapped ions~\cite{[viii]}.} 

We note that $H$ is given in units of a characteristic oscillator energy $\hbar \omega$ and time $t$ is rendered dimensionless with the same characteristic frequency $\omega$. In \Fig{fig:pot} we show the (non-linear) potential of \Eq{eq:Ham} as a dotted line. 
Intuitively, one can associate the dashed and solid lines in \Fig{fig:pot} with an effective oscillator potential associated with the coupling \Eq{eq:Hint} and the qubit being in the ground state \ket{g} (with $\EX{\sigma_z}=-1$) or excited state \ket{e} (with $\EX{\sigma_z}=1$) respectively. So we associate with the ground state of the qubit a quartic potential and with the excited state the potential energy of the Duffing oscillator. 

In Eq.~(\ref{eq:Ham}) $\beta$ is a dimensionless parameter representing
a scaling of the classical action of the oscillator with
respect to a Planck cell and $g$ is the strength of the applied
driving term. In this work we have chosen $g = 0.3$
and $\beta = 0.1$. For the Duffing oscillator, these have been
shown to produce Poincar\'e sections in good agreement
with solutions to the equivalent classical equations of motion~\cite{Per98,Brun:1996p393}, and so for these parameters the measurement device
is recognised to be classical. We have chosen such a system deliberately,
as the manifestation of a chaotic-like trajectory is a very
clear signature of the device operating on the classical side of the
quantum to classical transition.
Obviously other (non-chaotic) systems could equally
well serve as a measurement device; however for these it must
be clear that the device is operating on the classical side of the
quantum to classical transition, otherwise one has to proceed
further up the ``von Neumann chain'' of systems to find a degree
of freedom that is manifestly classical and thus capable of recording
the measurement outcome.
Our choice of a
non-linear oscillator can be used to clearly demonstrate
that a classical-like record has been achieved, without requiring
further verification or the need to move further up the chain and consider
the systems that measures or observes the oscillator. We note that whilst smaller
values of $\beta$ would produce even better matches with classical
dynamics, some of the results presented in this work
are computationally very demanding and set a practical
limit for us of $\beta= 0.1$ at this time. We note that a similar system
was recently used to explore the decoherence of qubits
coupled to a quantised Duffing oscillator~\cite{Buric:2009p022101}.

In order to suppress quantum interference effects we couple the measurement device to an environment using a Lindblad~\cite{Lindblad:1976p1778} type master equation. Here, the evolution of the
reduced density operator, $\rho(t)$, of the quantum system of interest is found by tracing out the environmental degrees of freedom that are responsible for dissipation. For our purposes, this environment---a bath of quantum degrees of freedom---is assumed to be Markovian. That is, there are no memory effects and the evolution of $\rho(t)$ depends only on $\rho(t)$, the Hamiltonian and operators representing interaction with the environment, termed Lindblad operators. The (non-unitary) dynamics of this (reduced) density operator is then given by:
\bel{lindblad}
\dot{\rho}=-i
\left[H,\rho\right] + \sum_m \left(L_m \rho L_{m}^{\dagger}
- \frac{1}{2}L_{m}^{\dagger} L_m \rho -\frac{1}{2} \rho L_{m}^{\dagger} L_m \right) .
\ee
Modelling a zero temperature dissipative environment requires just one Lindblad operator, $L=\sqrt{2\Gamma}a$ where $a$ is the oscillator annihilation operator. 
\NEW{It is possible to model finite temperature using a second Lindblad proportional to the creation operator, with suitable temperature-dependent coefficients, allowing for stimulated excitation and emission in addition to the spontaneous emission that represents zero temperature dissipation. However, we do not explore finite temperature here. }
Subjecting an oscillator to dissipation also requires the addition of a term ${\Gamma}\left(pq+qp \right)/2$ to the Hamiltonian \Eq{eq:Ham}. This term  is required to ensure that the (dissipative) classical-like dynamics of the oscillator term are correctly recovered in the correspondence limit~\cite{Brun:1996p393, Per98}. For a linear oscillator the term produces the expected frequency shift. So that our effective Hamiltonian is:
\ba
H&=&\frac{3}{4}p^2+\frac{\beta^2}{4}q^4-\frac{1}{4}q^2+\frac{g}{\beta}\cos(t)q-\nonumber \\
&&\frac{1}{4}\left(p^2+q^2\right)\sigma_z+\frac{\Gamma}{2}\left(pq+qp \right)
\label{eq:Ham}
\ea

\section{Measurement in the ensemble average \label{EA}}

Let us now consider how the qubit and oscillator system evolves under \Eq{lindblad}, using \Eq{eq:Ham} with the dissipative addition. We employ dissipation set by $\Gamma=0.125$ and an initial pure quantum state of $\frac{1}{\sqrt{2}}\left(\ket{g}+\ket{e}\right)\otimes\ket{\alpha \approx 6.8}$. Here \ket{\alpha} denotes a coherent state with average photon, or excitation, number $|\alpha|^2$ and is defined as usual by:
$$
\ket{\alpha}=e^{-|\alpha|^{2}/2} \sum_{n=0}^{\infty} \frac{\alpha^n}{\sqrt{n!}}\ket{n}.
\label{coh}
$$
We note that \ket{\alpha \approx 6.8} corresponds in phase space to a Gaussian bell centred at $(q=11,p=0)$. This initial oscillator state is chosen so as to minimize transient behaviour; however other choices equally well enable the possibility of either measurement result for the qubit $z$-value. 
\NEW{Clearly in a ``good'' quantum measurement experiment, the apparatus should not be primed so as to exclude the possibility of registering certain possible outcomes.}

\begin{figure}[tb]
\begin{center}
\resizebox*{0.4\textwidth}{!}{\includegraphics{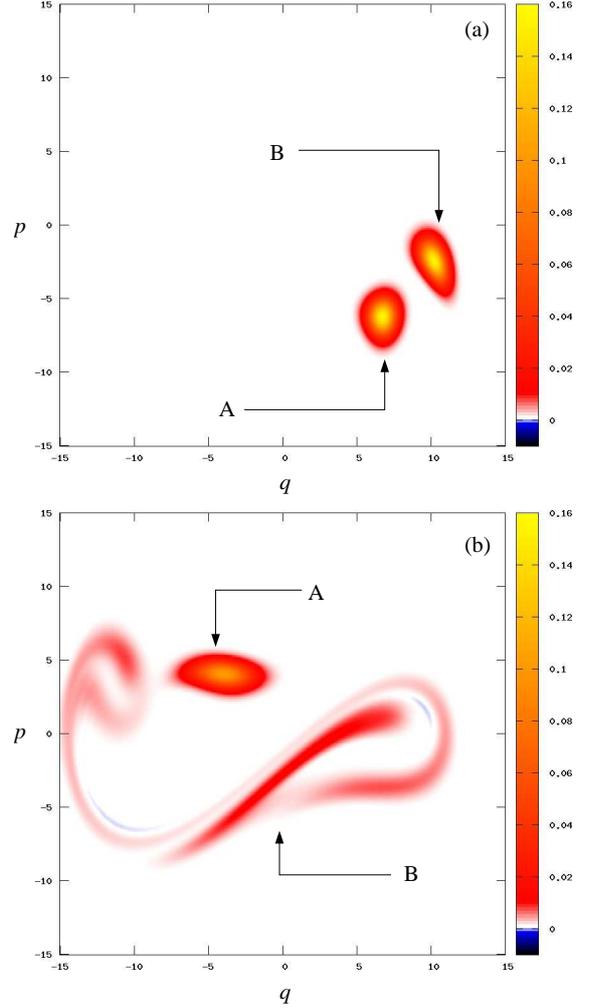}}
\end{center} \caption{(colour online) Wigner function of the field for solutions to the master equation where the initial density operator corresponds to the pure state $\frac{1}{\sqrt{2}}\left(\ket{g}+\ket{e}\right)\otimes\ket{\alpha \approx 6.8}$. (a) at $t/2\pi=0.1$ and (b) at $t/2\pi=2.25$. Here region A and B corresponds to measuring the qubit in state \ket{g} and \ket{e} respectively.}
\label{fig:wig}
\end{figure}

\NEW{The Wigner function, which can be viewed as a pseudo probability density function in the $(q,p)$ phase space, if very useful in gaining an good insight into the behaviour of dynamical systems. It can be defined by:
$$
W(q,p)=\frac{1}{2\pi}\int d\zeta\, \IPR{q+\frac{\zeta}{2}}{\rho}{q-\frac{\zeta}{2}}\exp{(-i\zeta p)}
$$
For a detailed discussion see, for example,~\cite{Schleich2001}. In \Fig{fig:wig} we show two snapshots of the Wigner function for the oscillator mode -- having traced out the qubit (see the on-line animation that accompanies this letter for a movie of the evolution of Wigner function the snapshot shown in \Fig{fig:wig}).} \Fig{fig:wig}(a) is taken very early on in the system evolution at $t/2\pi=0.1$. Here we see that the effect of the qubit on the measurement device is to split the coherent state up into two coherent state-like lumps. This is not surprising, as a very similar effect is seen in collapse and revival phenomena of the Jaynes-Cummings model of a qubit interacting with a harmonic oscillator. However, as the system evolves we see in \Fig{fig:wig}(b) ($t/2\pi=2.25$) the distribution becomes somewhat more interesting. One lump is still quite small and is associated with the qubit's ground state. The second is much larger, arising from the chaotic-like behaviour of the Duffing oscillator and is associated with the qubit's excited state.

\begin{figure}[tb]
\begin{center}
\resizebox*{0.4\textwidth}{!}{\includegraphics{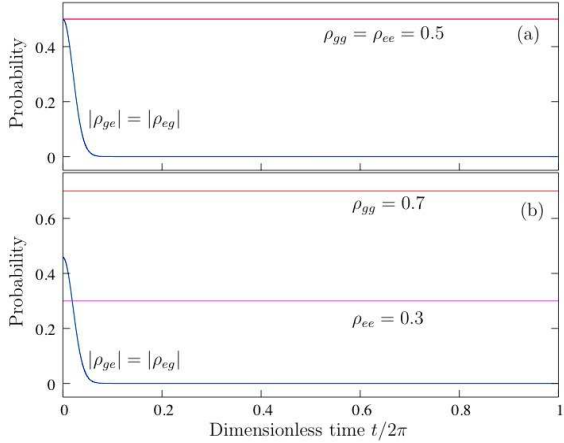}}
\end{center} \caption{(colour online) Elements of the reduced density matrix for the qubit found by solving~\Eq{lindblad} over one period of the external drive field. In (a) the initial state was $\frac{1}{\sqrt{2}}\left(\ket{g}+\ket{e}\right)\otimes\ket{\alpha \approx 6.8}$. Further confirmation of 
the Born rule is shown in (b) with the same matrix elements for the initial state $\left(\sqrt{0.7}\ket{g}+\sqrt{0.3}\ket{e}\right)\otimes\ket{\alpha \approx 6.8}$.}
\label{fig:probs}
\end{figure}Let us now turn to the qubit and consider its reduced density matrix
\bel{denOp}
\rho_Q:=\mathrm{Tr}_\mathrm{field}({\rho}):=\left(\begin{array}{cc}
 \rho_{gg}   & \rho_{ge}   \\
  \rho_{eg}  & \rho_{ee} 
\end{array}\right).
\ee
If we are indeed modelling a measurement process we would expect that, as the system evolves, $\rho_Q$ will diagonalise indicating that it has become a statistical mixture of the ground and excited state. \NEW{Such behaviour is clearly seen in the results presented, for the initial state
$$\ket{\psi(0)}=\frac{1}{\sqrt{2}}\left(\ket{g}+\ket{e}\right)\otimes\ket{\alpha \approx 6.8}$$ 
in \Fig{fig:probs}(a). Our measurement apparatus and interaction are arranged to keep $\langle \sigma_z \rangle$ fixed in time and thus yield the Born rule. Our choice of initial state as an equal superposition of \ket{g} and \ket{e} is rather special in that it is symmetric. In order to ensure that our results are not just some special case associate with this symmetric superposition, and the the Born rule holds, we show in \Fig{fig:probs}(b) the same data but for the initial state 
$$\ket{\psi(0)}=\left(\sqrt{0.7}\ket{g}+\sqrt{0.3}\ket{e}\right)\otimes\ket{\alpha \approx 6.8}.$$}

\begin{figure}[tb]
\begin{center}
\resizebox*{0.4\textwidth}{!}{\includegraphics{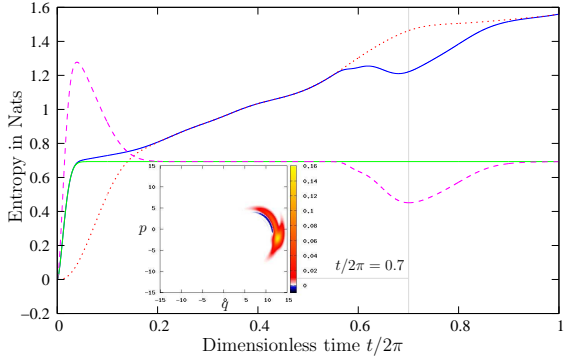}}
\end{center} \caption{(colour online) The time evolution of the entropy of the qubit $S_Q$ is shown in light grey (green), the oscillator $S_O$ in dark grey (blue), the total system $S$ as a dotted line (red) and the index of correlation $S_Q + S_O - S$ as a dashed line (magenta). The inset shows the Wigner function for the oscillator at $t/2\pi = 0.7$.
}
\label{fig:ents}
\end{figure}

Although the expected diagonalisation of $\rho_Q$ must happen in a measurement, this is clearly not a sufficient signature. The measurement apparatus needs to demonstrate an outcome that correlates to the measured state of (in our case) the qubit. To see that this is happening we examine the various entropies defined through
$$S\left(\rho\right)=-\mathrm{Tr}\left( \rho \ln \rho\right)$$ in \Fig{fig:ents}. It can be seen that the initial stage of the measurement involves entanglement of the qubit and the oscillator, as the two entropies $S_Q$ and $S_O$ rise rapidly to $\ln 2$ whilst the total entropy $S$ remains very close to zero. 

The measurement can be deemed to be completed when the oscillator and total entropies converge (and so the index of correlation $I=S_Q + S_O - S$ settles to the qubit entropy value of $\ln 2$), at around $t/2\pi = 0.2$. We note two further points about this figure. First, in general the oscillator (and thus total) entropy continues to climb after the measurement is completed. This is due to the continuing evolution of the oscillator, which has a chaotic signature for one of the two measurement outcomes. Second, there is actually a divergence between $S_O$ and $S$ located around $t/2\pi = 0.7$. This is nothing to do with quantum erasure or anything like that. It is simply due to loss of classical information. If we only look at our measurement apparatus around this time, the signatures corresponding to the two measurement outcomes are not completely distinct. To illustrate this, \Fig{fig:ents} contains an inset of the Wigner function for the oscillator at this time. \NEWT{We note that the classical records become distinct again after $t/2\pi\approx 0.9$ until the Wigner functions overlap once more -- at the same point in every drive cycle}.

\section{Single Measurements\label{SM}}

The master equation is very useful insofar as it tells us about the behaviour of the ensemble averages. However, we wish to gain some understanding about the way in which a single measurement (or classical record) can emerge as a natural process. Hence, we need to model the behaviour of individual realisations of the master equation. We therefore unravel the pure state evolution equation for the
system, that is equivalent to \Eq{lindblad} in the ensemble average over many realisations. The unravelling we use here is quantum state diffusion (QSD)~\cite{Gisin:1993p957,Gisin:1993p918,Per98}. For QSD, the state evolves according to
\ba
\ket{d\psi} &=& -i
H \ket{\psi} dt
+ \sum_m \left(L_m - \langle L_m \rangle_{\psi} \right) \ket{\psi} d\xi_m
+ \nonumber \\&& 
\sum_m \left(\langle L_{m}^{\dagger} \rangle_{\psi} L_m - \frac{1}{2} L_{m}^{\dagger} L_m
- \frac{1}{2} \langle L_{m}^{\dagger} \rangle_{\psi} \langle L_{m} \rangle_{\psi} \right)
\ket{\psi} dt\nonumber \\&& 
\label{qsd}
\ea
where the operators are defined as in \Eq{lindblad} and the $d\xi_m$ are independent complex differential random variables satisfying
$\overline{d\xi_m}=\overline{d\xi_m d\xi_n}=0$ and $\overline{d\xi_m^* d\xi_n}= dt\,\delta_{mn}$ (where the over-line denotes mean over an ensemble). The evolution of \Eq{qsd} is equivalent to that of \Eq{lindblad} when the ensemble mean of $\rho=\overline{\OP{\psi}{\psi}}$ over the noise is taken. Of course, one issue with unravelling a master equation is that in general there is no unique unravelling---many different approaches exist. However, in the specific scenario we consider here, the non-linear oscillator is made as classical as our computations permit. In this classical limit the distinction between unravellings begins to go away, in the sense that any unravelling of a dissipative and classically chaotic system must start to show a characteristic of the appropriate classical chaotic trajectory.
\begin{figure}[tb]
\begin{center}
\resizebox*{0.4\textwidth}{!}{\includegraphics{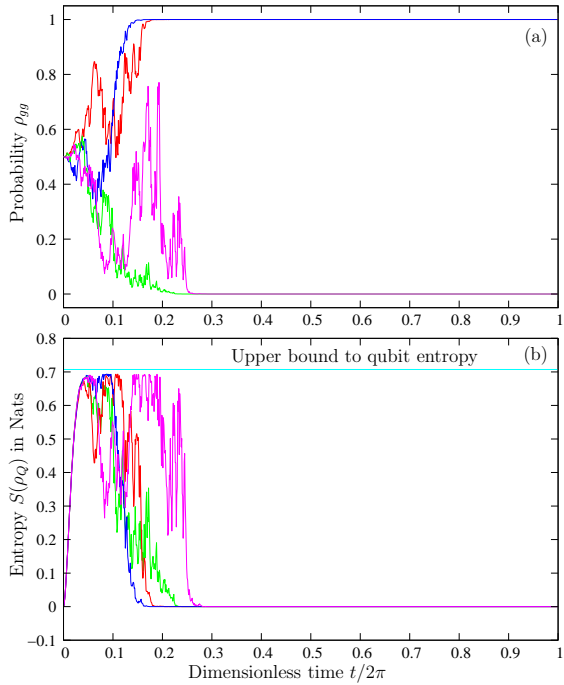}}
\end{center} 
\caption{(colour online) (a) Probability of the qubit being in \ket{g}
for four possible realisations of the mater equation. Each trajectory was determined using 
\Eq{qsd} with initial state $\frac{1}{\sqrt{2}}\left(\ket{g}+\ket{e}\right)\otimes\ket{\alpha \approx 6.8}$ with different random number seeds being used to generate each realisation. (b) The entropy of the qubit associated with each trajectory.
}
\label{fig:traj}
\end{figure}
In \Fig{fig:probs} we plotted, amongst other quantities, the ensemble average probability of being in the ground state. In \Fig{fig:traj}(a) we show the probability of being in the ground state for four solutions of~\Eq{qsd}. We clearly see that in each run the qubit is quickly projected into either \ket{g} or \ket{e}. We would expect that for these dynamics to be modelling a measurement, the oscillator initially entangles with the qubit (the initial stage of the measurement) and then disentangles (once the measurement has been completed). In \Fig{fig:traj}(b) we plot the entanglement entropy $$S\left(\rho_Q\right)=-\mathrm{Tr}\left( \rho_Q \ln \rho_Q\right)$$ for each trajectory. We have examined many more individual realisations of the master equation than shown here and this behaviour is reflected across the ensemble.

\begin{figure}[tb]
\begin{center}
\resizebox*{0.4\textwidth}{!}{\includegraphics{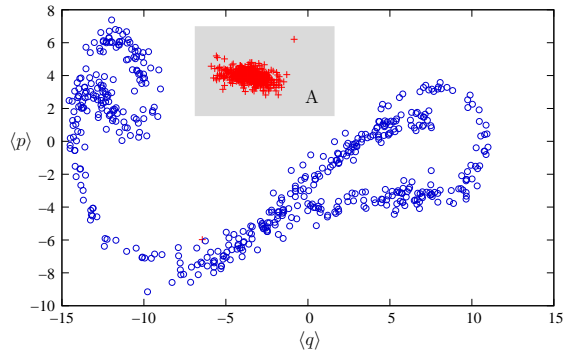}}
\end{center} \caption{(colour online) Example Poincar\'e sections, taken at $t/2\pi=n+1/4, n\in\mathbb{N}$, found by solving~\Eq{lindblad}. Circles denote a section where the qubit state \ket{e} is measured and crosses for \ket{g}. \NEW{Note that, excluding the first data point, all points of the periodic attractor measurement of \ket{g} lie within the approximate region A and are distinct from the chaotic like attractor corresponding to the qubit being in \ket{e}.}
}
\label{fig:ps}
\end{figure}
We now have all the ingredients of a measurement process apart from the most important one. We require that for each individual run a distinct classical record of the measured qubit value (corresponding to the state \ket{e} or \ket{g} to which the qubit projects) be present in the oscillator dynamics. We have seen in \Fig{fig:traj}(a) that the qubit projects into either \ket{e} or \ket{g} in each run and we recall that in \Fig{fig:pot} a different effective potential for the oscillator is associated with each of these states. In \Fig{fig:ps} we demonstrate how this translates into a dynamical record of the quantum measurement. Here we display two Poincar\'e sections, which arise with probabilities set by the Born rule. If the qubit is projected to \ket{g} then the dynamics become periodic and we obtain the attractor shown with crosses. Here, all but the first data point (due to transients) lie in the approximate region A. On the other hand, if the qubit is projected to the state \ket{e} the effective potential is that of the Duffing oscillator and we obtain a chaotic-like attractor (shown with circular points). The emergence of a chaotic-like solution gives a strong indication that the oscillator is sufficiently classical to perform a good---correspondence limit---model of measurement apparatus. We note that, while it takes a long time for the Poincar\'e section for a chaotic attractor to build up in order to measure \ket{e} it would be sufficient to demonstrate ``not in A'' after only a few drive periods. 
It may therefore be possible to perform the measurement on
a much shorter time scale by examining the individual
phase portraits of the oscillator and establishing some
kind of clear classification criteria. In principle it ought
to be possible to determine the measurement outcome by
around time $t=2/\pi = 0.2$, as shown in the entropy plot
of \Fig{fig:ents}. As stressed earlier, the chaotic trajectory is not
a requirement for any measurement device, but the ability to
produce it here demonstrates that the device is sufficiently classical
to record the measurement outcome, without the need to
then consider what observes this device.


\section{Incompatible Observables}
In the preceding discussion we have assumed that the qubit has been arranged so that there is no direct coupling between its possible states. It is natural to next consider what would happen if this was not the case. Coupling between \ket{g} and \ket{e} will arise if we add to the Hamiltonian~\Eq{eq:Ham} any operator that does not commute with $\sigma_z$. As an example we choose to use $\sigma_x$. In order to ensure that this addition will produce a significant effect we make it comparable in magnitude to the cross-Kerr coupling term. Hence, we choose to consider a new Hamiltonian of the form:
\ba
H&=&\frac{3}{4}p^2+\frac{\beta^2}{4}q^4-\frac{1}{4}q^2+\frac{g}{\beta}\cos(t)q+\frac{\Gamma}{2}\left(pq+qp \right)\nonumber \\
&&-\frac{1}{4}\left(p^2+q^2\right)\sigma_z+\frac{\sigma_x}{2}\label{eq:HamIO}
\ea
where we have added $\sigma_x/2$ to~\Eq{eq:Ham}.

In section~\ref{EA} we verified that the Born rule held in the ensemble average by solving~\Eq{lindblad} for the system and computing the reduced density matrix according to~\Eq{denOp}. Let us consider this situation and again solve~\Eq{lindblad} but now using the Hamiltonian of ~\Eq{eq:HamIO}.
Because we have introduced $\sigma_x/2$ to the Hamiltonian the dynamics will not be so trivial. Once more we choose the initial state of the system to be $\frac{1}{\sqrt{2}}\left(\ket{g}+\ket{e}\right)\otimes\ket{\alpha \approx 6.8}$.

\begin{figure}[!tb]
\begin{center}
\resizebox*{0.4\textwidth}{!}{\includegraphics{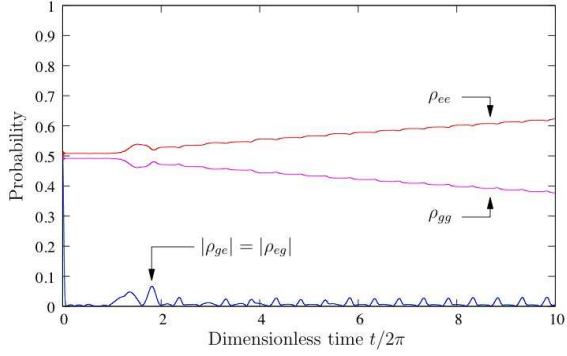}}
\end{center} 
\caption{(colour online) For comparison with~\Fig{fig:probs}(a)  (but over a much larger duration). Elements of the reduced density matrix for the qubit found by solving~\Eq{lindblad} over one period of the external drive field but now using the Hamiltonian~\Eq{eq:HamIO}}. 
\label{fig:probs2}
\end{figure}
\begin{figure}[!tb]
\begin{center}
\resizebox*{0.4\textwidth}{!}{\includegraphics{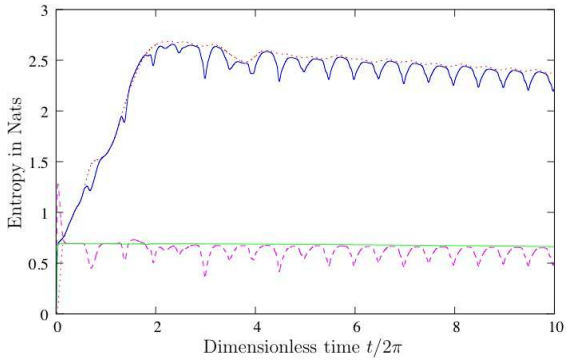}}
\end{center} \caption{(colour online) For comparison with~\Fig{fig:ents} (but over a much larger duration). The time evolution of the entropy of the qubit  found by solving~\Eq{lindblad} over one period of the external drive field but now using the Hamiltonian~\Eq{eq:HamIO}. $S_Q$ is shown in light grey (green), the oscillator $S_O$ in dark grey (blue), the total system $S$ as a dotted line (red) and the index of correlation $S_Q + S_O - S$ as a dashed line (magenta).
}
\label{fig:ents2}
\end{figure}

In~\Fig{fig:probs2} we show the dynamics of the elements of $\rho_Q$ for comparison with~\Fig{fig:probs}(a). In order to get a good idea of the long term effect of introducing  $\sigma_x/2$ to the Hamiltonian, this figure has been computed over a much longer time scale. Here we see that, although not perfect, there is very good agreement between the two models and the Born rule is closely adhered to until $t/2\pi \approx 1$. From our previous arguments we observe that this may well be sufficient evolution so as to be able to determine the state of the qubit from the dynamics of the oscillator component. In which case, our set up of using a non-linear oscillator in its correspondence limit to record the state of the qubit is still capable of making a good projective measurement. However, as $t/2\pi$ increases beyond \NEWT{unity} there is an increasing deviation from the Born rule. 

For comparison with~\Fig{fig:ents} we show in~\Fig{fig:ents2} the entropic quantities of $S$, $S_O$, $S_Q$ and the index of correlation. Although it is not obvious from this figure we note that the qubit entropy, although initially rapidly ascending to its maximum value of $\ln 2$, decreases slightly over time.

\begin{figure}[!t]
\begin{center}
\resizebox*{0.4\textwidth}{!}{\includegraphics{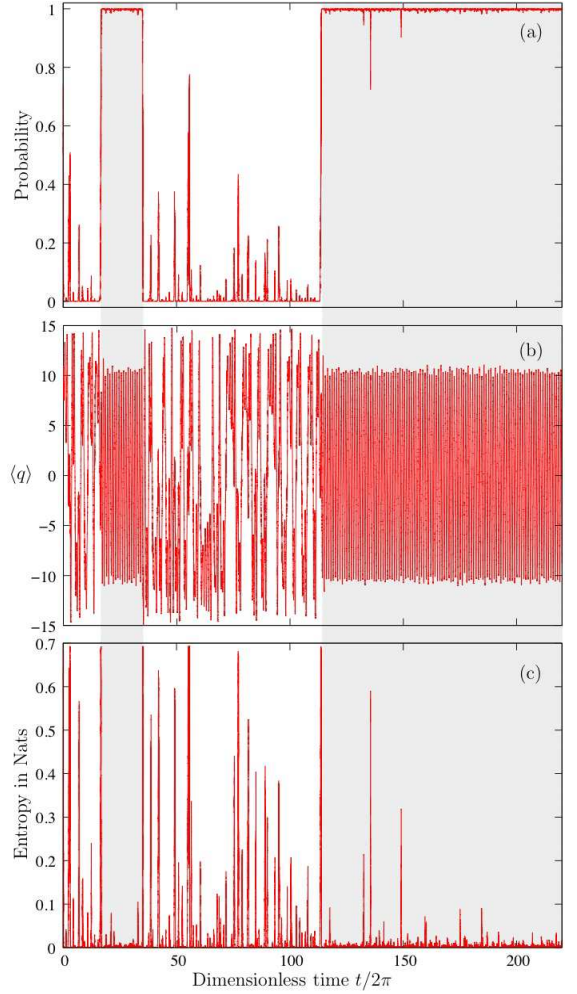}}
\end{center} 
\caption{(colour online) One possible realisations of the mater equation. Each trajectory was determined using 
\Eq{qsd} with initial state $\frac{1}{\sqrt{2}}\left(\ket{g}+\ket{e}\right)\otimes\ket{\alpha \approx 6.8}$ but now using the Hamiltonian~\Eq{eq:HamIO} (a) Probability of the qubit being in \ket{g} (b) expectation value of position \EX{q} (c) Entanglement entropy in Nats. Here the grey region indicates the points where the qubit is - or very near - to the state \ket{g} and has been added to aid comparison between (a), (b) and (c).
}
\label{fig:traj2}
\end{figure}

We have seen that introducing $\sigma_x/2$ results in the Born rule being not well observed beyond $t/2\pi \approx 1$ in the solution to the master equation. However, in analogy with section~\ref{SM}, let us now not consider a statistical ensemble and ask what happens in an individual experiment. Once more we use the quantum state diffusion unravelling of the master equation as given in~\Eq{qsd}. Recall that it is via the introduction of the environment that we localise the state vector and can recover classical-like trajectories. Furthermore, in unravelling the master equation we introduce some stochastic behaviour that represents fluctuations due to environmental degrees of freedom. The oscillator is effected by a Brownian motion and the $\sigma_x/2$ term couples the eigenstates of the qubit together.  We might therefore expect, that taking these two factors together with the cross-Kerr coupling into account, to see some kind of switching behaviour in the qubit. In~\Fig{fig:traj2}(a) shows a typical trajectory and we see that it does indeed operate in this manner.

In our previous discussion and with reference to~\Fig{fig:ps} we established that if the qubit is projected to \ket{g} or \ket{e} then the dynamics of the expectation values of the oscillator become either periodic or chaotic-like respectivly. In~\Fig{fig:traj2}(b) we plot \EX{q} as a function of time and by comparison with~\Fig{fig:traj2}(a) we observe that the same behaviour is still manifest despite the introduction of $\sigma_x/2$ into the Hamiltonian.

That is, the oscillator closely monitors the state of the qubit, exhibiting periodic or chaotic-like oscillations when the qubit is in, or near to, \ket{g} or \ket{e} respectively. We know from our analysis of the master equation, that this is not a \emph{``good''} measurement within the Born framework. Nevertheless, the classical like dynamics of the oscillator provide a clear record of the state of the qubit. In this sense we could consider that effective measurements are still being made. In addition, as we are generating a classical like continuous monitor of the quantum state of the qubit it may well be that quantum circuits constructed in this way may find utility in quantum feedback and control problems (see, for example, \cite{Ralph:2004p100}).

In~\Fig{fig:traj2}(a) we observe a certain spikiness in the probability amplitude. An explanation for this can be found if we consider the entanglement entropy which we show in ~\Fig{fig:traj2}(c). Here we see that these spikes in probability are accompanied by concomitant spikes in the entropy. So, the $\sigma_x$ term in the Hamiltonian is trying to take the qubit out of an eigenstate of $\sigma_z$. But the qubit is coupled to the oscillator mode via $\left(p^2+q^2\right)\sigma_z/4$. So accompanying this move to a superposition of the qubit is a tendency for the qubit and the oscillator to entangle. Now, the oscillator is subject to environmental decoherence which will try to suppress this entanglement. Hence, we can understand this spiky behaviour in terms of a competition between these effects. Furthermore we can see that it is when the entanglement is very large that the qubit is most likely to switch between \ket{g} and \ket{e}. So that this quantum jumping behaviour between the eigenstates is a consequence of trying to projectively measure the qubit into an eigenstate of an observable that is incompatible with the qubit Hamiltonian.

\begin{figure}[!t]
\begin{center}
\resizebox*{0.4\textwidth}{!}{\includegraphics{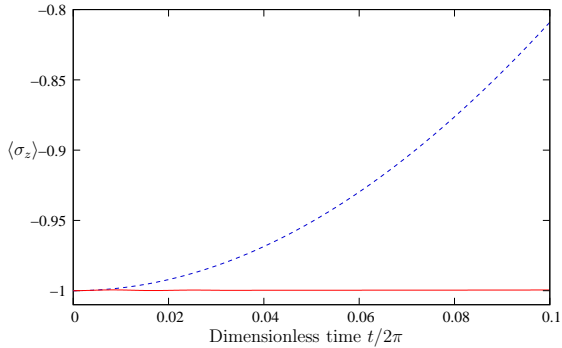}}
\end{center} 
\caption{\NEWT{(colour online) Illustration of the Zeno effect for the system in initial state \ket{g}. Here  \EX{\sigma_z} is plotted as a function of dimensionless time for (solid/red line) qubit subjected to a measurement of $\sigma_z$ (dashed/blue line) free evolution of the qubit.}}
\label{fig:zeno}
\end{figure}

\NEWT{Further illustration of the measurement effect of the oscillator
   on the qubit can be shown through the Zeno suppression of
   coherent oscillation. If the qubit with Hamiltonian as in \Eq{eq:HamIO} is
   alternatively started in state \ket{g}, we know that with no coupling
   to the oscillator it exhibits coherent oscillation, for example in
   \EX{\sigma_z} as a function of time. If subjected to measurement of
   $\sigma_z$, the expectation is that the (initially quadratic)
   evolution in \EX{\sigma_z} would be retarded through the Zeno effect.
   \Fig{fig:zeno} illustrates a very strong Zeno suppression of the
   qubit evolution for our system when the oscillator couples to it, consistent with a strong measurement of $\sigma_z$.}

\section{Conclusion}

In this letter we have presented a fully quantum mechanical
model of a projective measurement process. The
measurement device comprises an oscillator circuit with
a dissipative environment, where the dynamics of expectation
values, in the correspondence limit, are either
chaotic-like or periodic depending on the measured value
and projected state of a qubit. The device parameters are thus
chosen so it is clear that this device is on the classical side of
its quantum to classical transition in behaviour. Its ``record'' of
the measurement outcome is therefore classical, and there is
no need to go further up the chain and consider
what system might be used to observe this device.
In our model no preferred
basis was assumed to exist \emph{a priori}, rather it emerged
from the coupling between the measurement device and
the quantum object. In ensemble language our measurement
device and qubit attained the expected final
mixtures. However, we have further demonstrated that
individual classical-like trajectories of an open quantum
system can act as a record of the measurement of an individual
qubit, in line with the Born rule. Even for the case of chaotic
apparatus it is possible to produce decent measurements.

\bibliographystyle{elsarticle-num}
\bibliography{ref}
\end{document}